\documentclass{PoS}

\title{Diboson Production at D0}

\ShortTitle{Diboson Production at D0}

\author{\speaker{Joseph HALEY}%
         \thanks{For the D0 Collaboration.}\\
        Princeton University\\
        E-mail: \email{cooljoe@princeton.edu}}

\abstract{ We present recent diboson production measurements from the
  D0 experiment at Fermilab's Tevatron collider. The production of
  $ZZ$ was observed using leptonic final
  states. $Z\gamma\to\nu\nu\gamma$ was observed and used to set the
  most stringent limits from a hadron collider on anomalous
  $Z\gamma\gamma$ and $ZZ\gamma$ trilinear gauge couplings
  (TGCs). $WW$ events with leptonic final states and $WW+WZ$ events
  with semi-leptonic final states were used to set limits on anomalous
  $WWZ$ and $WW\gamma$ TGCs. Finally, limits on anomalous $WWZ$ and
  $WW\gamma$ TGCs were obtained from a combination of the
  fully-leptonic $W\gamma$, $WW$, and $WZ$ channels and the
  semi-leptonic $WW$ and $WZ$ channels, giving the most stringent
  limits from a hadron collider.  }

\FullConference{European Physical Society Europhysics Conference on High Energy Physics\\
		 July 16-22, 2009\\
		 Krakow, Poland}

\def\met{\ensuremath{E\kern-0.57em/_{T}}}
\begin{document}

\section{Introduction}

Diboson productions provides a probe for new physics that may reside
beyond some high energy scale $\Lambda_{NP}$.  In this case, the
Standard Model (SM) of particle physics is simply the low energy limit
of a more general theory.  Such new physics could result in anomalous
trilinear gauge-boson couplings (TGCs) that would affect the
production rates and kinematics of diboson processes.  Measuring TGCs
could give clues to the new physics and the mechanism for electroweak
symmetry breaking.

The most general Lorentz-invariant Lagrangian for $\gamma WW$ and
$ZWW$ TGCs has 14 coupling parameters.  However, the theory simplifies
to a more manageable five coupling parameters by assuming EM gauge
invariance and C, P, and CP symmetry conservation.  These five
couplings, $g_1^Z$, $\kappa_\gamma$, $\kappa_Z$, $\lambda_\gamma$, and
$\lambda_Z$, are measured by analyzing the production of $WW$, $WZ$,
and $W\gamma$.  $\gamma ZZ$ and $\gamma\gamma Z$ TGCs are probed by
analyzing $ZZ$ and $Z\gamma$ production.  For $ZZ\gamma$ and
$Z\gamma\gamma$ TGCs in $Z\gamma$ production we consider the four
coupling parameters $h_3^\gamma$, $h_3^Z$, $h_4^\gamma$, and $h_4^Z$
from the most general Lorentz-invariant Lagrangian that conserves CP
symmetry.  In the SM, $g_1^Z=\kappa_\gamma=\kappa_Z=1$ and
$\lambda_\gamma=\lambda_Z=h_3^\gamma=h_3^Z=h_4^\gamma=h_4^Z=0$ with
any deviation defined as an anomalous TGC.

\section{$ZZ\to\ell\ell\ell\ell$}

The $ZZ\to\ell\ell\ell\ell$ analysis~\cite{bib:ZZllll} selected events
in 1.7~fb$^{-1}$ of D0 data with four energetic leptons
($e^+e^-e^+e^-$, $\mu^+\mu^-\mu^+\mu^-$, or $e^+e^-\mu^+\mu^-$).  The
leptons were grouped into oppositely changed, same flavor pairs
required to have an invariant dilepton mass $M_{\ell\ell}>70$~GeV for
one pair and $M_{\ell\ell}>50$~GeV for the other.  There was a small
background from $Z/\gamma^*$+jets events, in which two jets were
incorrectly reconstructed as electrons; however, there are no SM
backgrounds with four energetic leptons.  As a result, this final
state is extremely clean with only $0.14^{+0.03}_{-0.02}$ background
and $1.89\pm 0.08$ signal events expected.  Three events were observed
in the data resulting in a measured cross section of
$\sigma(ZZ)=1.75^{+1.27}_{-0.86}$(stat)$\pm0.13$(syst)~pb, which is
consistent with the SM NLO prediction of $\sigma(ZZ)=1.4\pm 0.1$~pb.
The observed significance of the measurement was 5.3$\sigma$ and
represents the first observation of $ZZ$ production at the Tevatron.

To further improve the measurement, this result was combined with the
$ZZ\to\ell\ell\ell\ell$ analysis using a previous dataset of
1~fb$^{-1}$ and the $ZZ\to\ell\ell\nu\nu$ analysis with 2.7~fb$^{-1}$
of data.  The combined measurement yields $\sigma(ZZ)=1.60\pm
0.63$(stat)$^{+0.16}_{-0.17}$(syst)~pb with an observed significance
of 5.7$\sigma$.

\section{$Z\gamma\to\nu\nu\gamma$}

The $Z\gamma\to\nu\nu\gamma$ analysis~\cite{bib:Zg} selected events in
3.6~fb$^{-1}$ of data with ``missing transverse energy'' $\met>70$~GeV
from the undetected neutrinos and a single high energy photon having a
``transverse energy'' of $E_T^\gamma>90$~GeV.  Backgrounds from
$W\to\ell\nu$ and $Z\to\ell\ell$ were reduced by vetoing events with
muons, additional EM objects, and isolated tracks.  Non-collision
backgrounds ({\it e.g.}, bremsstrahlung from beam halo) were reduced
by using the EM shower in the calorimeter to determine the origin of
the photon and require that it be consistent with coming from the
$p\bar{p}$ collision point.  Finally, mis-measured \met\ was minimized
by requiring zero jets.  With this selection, there were $17.3\pm 2.4$
background and $33.7\pm 3.4$ signal events expects and 51 candidate
events observed.  The measured cross section times branching ratio was
$\sigma(Z\gamma;
E_T^\gamma>90$~GeV)$\cdot$BR($Z\to\nu\nu)=32.9\pm9$(stat+syst)$\pm
2$(lumi)~fb, which is consistent with the SM NLO prediction of
$\sigma(Z\gamma; E_T^\gamma>90$~GeV)$\cdot$BR($Z\to\nu\nu)=39\pm4$~fb.
The observed significance of this measurement was $5.1\sigma$ and was
the first observation of $Z\gamma\to\nu\nu\gamma$ at the Tevatron.

The $E_T^\gamma$ spectrum is highly sensitive to the $\gamma ZZ$ and
$\gamma\gamma Z$ couplings and was used to set limits on anomalous
TGCs for $\Lambda_{NP}=1.5$~TeV.  The 95\% confidence level (CL)
limits are $|h_3^\gamma|<0.036$, $|h_3^Z|<0.0019$,
$|h_4^\gamma|<0.035$, $|h_4^Z|<0.0019$.  When combined with the
$Z\gamma\to ee\gamma$ and $Z\gamma\to\mu\mu\gamma$ channels using
1~fb$^{-1}$ of data, the 95\% CL limits become $|h_3^\gamma|<0.033$,
$|h_3^Z|<0.0017$, $|h_4^\gamma|<0.033$, $|h_4^Z|<0.0017$, which are
the world best for $h_3^Z$, $h_4^\gamma$, and $h_4^Z$.

\section{$WW/WZ\to\ell\nu q\bar{q}$}

The $WW/WZ\to\ell\nu q\bar{q}$ analysis~\cite{bib:WWWZ} selected
events in 1.1~fb$^{-1}$ of data with one high $p_T$ isolated lepton,
large \met\ (indicating a neutrino), and two high $p_T$ jets.
Background from multijet events, in which a jets was mistakenly
identified as a lepton, were reduced by requiring that the invariant
mass reconstructed from the lepton $\vec{p}_T^\ell$ and \met\ be
greater than 35~GeV.  $W$+jets and other backgrounds were separated
from the signal using a ``Random Forest'' multivariate discriminant.
The output distribution of the Random Forest discriminant was fit to
reveal a signal cross section of $\sigma(WW+WZ)=20.2\pm
4.4$(stat+syst)$\pm 1.2$(lumi)~pb, which is consistent with the SM NLO
prediction of $\sigma(WW+WZ)=16.1\pm 0.9$~fb.  The significance of
this measurement was $4.4\sigma$, representing the first evidence for
this process at the Tevatron.

The $p_T$ spectrum of the dijet system was used to set
limits~\cite{bib:WWWZ2} on anomalous $\gamma WW$ and $ZWW$ TGCs.  The
limits were measured for $\Lambda_{NP}=2$~TeV assuming two different
scenarios.  The first scenario, also used by the LEP experiments,
requires SU(2)$\times$U(1) symmetry.  This so-called LEP
parametrization requires $\lambda_\gamma=\lambda_Z$ and
$\Delta\kappa_Z=\Delta g_1^Z-\Delta\kappa_\gamma\tan(\theta_W)$; where
$\Delta\kappa_V\equiv\kappa_V-1$ and $\Delta g_1^Z\equiv g_1^Z-1$.
The 95\% CL limits on the three free parameters were measured to be
$-0.44<\Delta\kappa_\gamma<0.50$, $-0.10<\lambda<0.11$, and
$-0.12<\Delta g_1^Z<0.20$.  The second scenario assumes equal
couplings for $\gamma WW$ and $ZWW$ resulting in
$\lambda_\gamma=\lambda_Z$, $\kappa_Z=\kappa_\gamma$, and
$g_1^Z=g_1^\gamma=1$.  For the two free parameters in the equal
couplings scenario we measured 95\% CL limits of
$-0.16<\Delta\kappa<0.23$ and $-0.11<\lambda<0.11$.

\section{$WW\to\ell\nu\ell\nu$}

The $WW\to\ell\nu\ell\nu$ analysis~\cite{bib:WWlvlv} selected events
in 1.0~fb$^{-1}$ of data with two isolated high $p_T$ leptons of
opposite charge ($e^+e^-$, $e^\pm\mu^\mp$, or $\mu^+\mu^-$).  A cut on
\met\ was optimized for each channel to reduce $Z\to\ell\ell$
backgrounds.  $W$+jets and $t\bar{t}$ backgrounds were reduced by
requiring the $\vec{p}_T$ of the two leptons to be balanced with the
\met.  The cross section measured for $WW$ production was
$\sigma(WW)=11.5\pm 2.1$(stat+syst)$\pm 0.7$(lumi)~pb, consistent with
the NLO SM prediction of $\sigma(WW)=12.0\pm 0.7$~pb.

Limits on anomalous TGCs for both the LEP parametrization and equal
couplings scenario with $\Lambda_{NP}=2$~TeV were determined from the
two-dimensional distribution of the lepton $p_T$s.  In the LEP
parametrization, we measured 95\% CL limits of
$-0.54<\Delta\kappa_\gamma<0.83$, $-0.14<\lambda<0.18$, and
$-0.14<\Delta g_1^Z<0.30$.  For the equal couplings scenario, the
95\% CL limits were $-0.12<\Delta\kappa<0.35$ and
$-0.14<\lambda<0.18$.

\section{$WV$ Combination}

Four analyses with approximately 1~fb$^{-1}$ of data were combined to
improve the measurement of anomalous $\gamma WW$ and $ZWW$ TGCs.
Along with the $WW/WZ\to\ell\nu q\bar{q}$ and $WW\to\ell\nu\ell\nu$
analyses, the combination~\cite{bib:WVcombo} incorporated previous
analyses of $WZ\to\ell\nu\ell\ell$ and $W\gamma\to\ell\nu\gamma$.  The
combined measurement yielded 95\% CL limits for the LEP
parametrization of $-0.29<\Delta\kappa_\gamma<0.38$,
$-0.08<\lambda<0.08$, and $-0.07<\Delta g_1^Z<0.16$ with
$\Lambda_{NP}=2$~TeV.  For the equal couplings scenario with
$\Lambda_{NP}=2$~TeV, we measured 95\% CL limits of
$-0.11<\Delta\kappa<0.18$ and $-0.08<\lambda<0.08$.  These are the
most stringent limits from the Tevatron and are approaching the
sensitivity of the individual LEP2 experiments.

\section{Conclusions}

The recent diboson measurements from the D0 experiment represent many
first and best diboson measurements from the Tevatron.  $ZZ$
production was observed for the first time; $Z\gamma\to\nu\nu\gamma$
was observed and used to set the world best limits on $h_3^Z$,
$h_4^\gamma$, and $h_4^Z$; $WW\to\ell\nu\ell\nu$ and $WW/WZ\to\ell\nu
q\bar{q}$ were measured and used to set limits on anomalous TGCs; and
the best Tevatron limits on anomalous $WWZ$ and $WW\gamma$ TGCs were
obtained from a combination of four diboson analyses.  So far, all
diboson measurements from D0 are consistent with SM predictions.  The
future of diboson physics at D0 looks bright as we collect more data
each year and increase our sensitivity to new physics lurking beyond
the SM.

\end{document}